# Studies of Impurity-Doping Effects and NMR Measurements of La1111 and/or Nd 1111 Fe-Pnictide Superconductors

Masatoshi SATO,[1,2,*] Yoshiaki KOBAYASHI,[1,2] Sang Chul LEE,[1] Hidefumi TAKAHASHI,[1] Erika SATOMI,[1] and Yoko MIURA[1,2]

[1]*Department of Physics, Division of Material Science, Nagoya University, Furo-cho, Chikusa-ku, Nagoya 464-8602*
[2]*JST, TRIP, Nagoya University, Furo-cho, Chikusa-ku, Nagoya 464-8602*

Measurements of the electrical resistivities $\rho$, Hall coefficients $R_H$, thermoelectric powers $S$, electronic specific heat coefficients $\gamma$ have been carried out for samples of $LnFe_{1-y}M_yAsO_{1-x}F_x$ (Ln=La, Nd; M=Co, Mn; $x$=0.11) obtained by M atom dopings to the superconducting $LnFeAsO_{1-x}F_x$ (Ln1111) system. The NMR longitudinal relaxation rates $1/T_1$ have also been measured for samples of $LaFe_{1-y}Co_yAsO_{1-x}F_x$ with various $x$ values. Co atoms doped to the superconducting $LnFeAsO_{1-x}F_x$ are nonmagnetic, and the $T_c$-suppression rate $|dT_c/dx|$ by the Co atoms has been found to be too small to understand by the pair breaking effect expected for the $S_\pm$ superconducting symmetry proposed as the most probable one for the system. It throws a serious doubt whether the symmetry is realized in the system. Instead of the pair breaking, two mechanisms of the $T_c$-suppression by the doped impurities have been found: One is the electron localization, which appears when the sheet resistance $R_\Box$ exceeds $h/4e^2$=6.45 k$\Omega$, and another is the disappearance (or reduction) of the hole-Fermi-surfaces around the $\Gamma$ point in the reciprocal space. The latter mechanism has been observed, when the electron number increases with increasing Co-doping level and the system changes from an "anomalous metal" to an ordinary one. On the two distinct $T$ dependences of the NMR longitudinal relaxation rate $1/T_1$ of $LaFeAsO_{1-x}F_x$, $1/T_1 \propto T^6$ reported by our group in the $T$ region from $T_c$ to ~0.4 $T_c$ for samples with the highest $T_c$ values with varying $x$, and $1/T_1 \propto T^{2.5-3.0}$ observed by many groups in the almost entire $T$ region studied below $T_c$, we discuss what the origin of the difference is, and show that, at least, the $T^{2.5-3.0}$-like dependence of $1/T_1$ cannot be considered as the experimental evidence for the $S_\pm$ symmetry of $\Delta$.

**KEYWORDS:** $LaFe_{1-y}M_yAsO_{1-x}F_x$ (M=Co, Mn), NMR relaxation rate, $1/T_1$, transport properties, specific heats

## 1. Introduction

In the study of the origin of the high superconducting transition temperature $T_c$ of the newly found Fe-pnictide systems,[1] it is essential to identify the symmetry of their superconducting order parameter $\Delta$, and at the very early stage of its study, the $S_\pm$-symmetry has been proposed for the $LaFeAsO_{1-x}F_x$ (La1111) system as the most probable one[2,3] by considering that the superconductivity is realized by the spin fluctuations in the FeAs layers of edge-sharing FeAs$_4$ tetrahedra. It is characterized by the nodeless $\Delta$ on each Fermi surface but opposite signs of $\Delta$ values on the disconnected Fermi surfaces around $\Gamma$ and M points in the reciprocal space.

Various experiments have also been carried out to extract information on the symmetry of $\Delta$, and many groups have reported that $\Delta$ of La1111 is nodeless on each Fermi surface,[4-7] consistently with the $S_\pm$-symmetry. However, there is another important point, that is, we have to clarify whether the signs of the order parameters are opposite or not on the two disconnected Fermi surfaces around $\Gamma$ and M points. We have reported several papers and argued on this point based on our experimental studies carried out by means of transport[8-11] and NMR measurements on $LaFe_{1-y}Co_yAsO_{1-x}F_x$ ($x$=0.11).[12] One of important results obtained in the studies on the symmetry of $\Delta$ is that the rate of the $T_c$-suppression by the Co doping is much weaker than we expected for superconductors with node(s) of $\Delta$. It is also true even for superconductors without nodes on each Fermi surface but having opposite signs of the order parameters on disconnected Fermi surfaces.[8-11] Then, it is important to clarify whether the small $T_c$-suppression rate indicates that the superconducting order parameter of the system is different from the $S_\pm$ one. Another result of the studies is that the temperature ($T$) dependence of the NMR longitudinal relaxation rate $1/T_1$ observed by our group for La1111 samples with $x$=0.11,[12] can be approximately described by the relation $1/T_1 \propto T^6$ in the $T$ region between ~0.4$T_c$ and $T_c$, remarkably different from the relation $1/T_1 \propto T^{2.5-3.0}$ observed by many groups[13-17] in the wide $T$ region of (0.1~0.2)$T_c$<$T$<$T_c$ for samples including one with similar $T_c$ value to ours. Although the relation $1/T_1 \propto T^{2.5-3.0}$ has been explained by considering the $S_\pm$-symmetry of the order parameter,[18-20] it is difficult, as we have pointed out,[12] to understand the observations of these two distinct $T$ dependences simultaneously.

In this paper, we present detailed results on the small $T_c$ suppression rate obtained for samples of M-atom doped Ln1111 systems, $LnFe_{1-y}M_yAsO_{1-x}F_x$ (Ln=La, Nd; M=Co, Mn; $x$=0.11). We also present further results of the studies on the two distinct $T$ dependences of $1/T_1$, and argue whether the $S_\pm$-symmetry is realized in the present system or not.

## 2. Experiments

Polycrystalline samples of $LnFe_{1-y}M_yAsO_{1-x}F_x$ (Ln=La and Nd; M=Co and Mn and $x$ is nominal and fixed at 0.11 except the samples used in the NMR studies.) were prepared from initial mixtures of Ln, Ln$_2$O$_3$, LnF$_3$, FeAs and MAs with the nominal molar ratios.[9-11] The MAs powders were obtained by annealing mixtures of M and As in an evacuated quartz ampoules at 850 °C. (Hereafter, we use the nominal values of $x$ and $y$.) Details of the preparations and characterizations are in the previous papers:[9,10] The X-ray powder patterns were taken with Cuk$\alpha$ radiation for 7 s at each scattering angle 2$\theta$ at a step of 0.01°, where we found that the M-atom doping was successfully carried out (see the $y$ dependence of the lattice parameters shown in Figs. 1(a) and 2(a). The superconducting diamagnetic moments were measured by a Quantum Design SQUID magnetometer with the magnetic field $H$ of 10 G under both conditions of the zero-field cooling (ZFC) and field cooling (FC). From the data of the electrical resistivities ρ and diamagnetic moments, the $T_c$ values were determined as described previously.[9-11] These two kinds of $T_c$ values agree rather well for almost all the samples with the nominal $x$ of 0.11, but for the series of samples used in the studies of the NMR $1/T_1$ in the $x$ region $0.11 \leq x \leq 0.2$, we have found that the $T_c$ values determined by the resistivity measurements are slightly higher than those determined from the data of the superconducting diamagnetism. This difference between these two kinds of $T_c$'s seems to become more significant with increasing $x$.

*corresponding author (e43247a@nucc.cc.nagoya-u.ac.jp)

Hall coefficients $R_H$ of the polycrystalline samples were measured with increasing $T$ stepwise under the magnetic field $H$ of 7 T, where the sample plates were rotated around an axis perpendicular to the field, and thermoelectric powers $S$ were measured by the methods described in refs. 21 and 22. Measurements of the specific heats of the samples were carried out by a Quantum Design PPMS in the $T$ region between 2 and 60 K with increasing $T$ stepwise. The electronic specific heat coefficients γ of the samples with the superconducting transition were obtained by estimating the magnitudes of the specific heat jump $\Delta C$ at $T_c$ and simply assuming the relation $\Delta C = 1.43\gamma$. For other detailed data on the samples, see our previous papers.[9-11]

The $^{75}$As-NMR measurements were carried out for the samples of La1111 with various F concentration $x$ by the standard coherent pulse method,[12] where the spectra were measured by recording the integrated spin-echo intensity $I$ with the NMR frequency $f$ or applied magnetic field $H$ being changed stepwise. The spectra can be understood by using the quadrupole frequency $\nu_Q = 10.16$ MHz for $y=0$. In the study of the $^{75}$As-NMR $1/T_1$, the recovery curves were obtained by plotting the integrated intensity at a peak position of the $I-f$ curves ($f \cong 51.04$ MHz and $H=6.94$ T or $f \cong 44.26$ MHz and $H=6.005$ T) against the time $t$ elapsed after the saturation pulse. The peak corresponds to the condition of $H$ in the $ab$ plane. (The typical spectral shape taken for the same sample as used here can be found in Fig. 5 of ref. 9, where the peak is indicated by an arrow.)

## 3. Experimental Results
### 3.1 Superconducting transition temperatures and the electrical resistivities

In Fig. 1(a), the lattice parameters $a$ and $c$ are plotted for the M-atom doped La1111 system (LaFe$_{1-y}$M$_y$AsO$_{1-x}$F$_x$) against $y(-y)$ for M=Co(Mn), and the $T$ dependences of the electrical resistivities ρ of the Mn- and Co-doped samples are shown in Figs. 1(b) and 1(c), respectively, where the $y$ value is attached to the corresponding data. Similar figures are shown in Figs. 2(a)-2(c) for the M-atom doped Nd1111 system NdFe$_{1-y}$M$_y$AsO$_{1-x}$F$_x$ (M=Co and Mn and $x$=0.11). From the data of the lattice parameters, we can find that the M-atom doping was successful. It can be also said that $y$ does not deviate significantly from the nominal value, because the parameter $c$ is linear in the entire $y$ region between 0.0 and 1.0, at least for LaFe$_{1-y}$Mn$_y$AsO$_{1-x}$F$_x$ (see Fig. 1(a)).

The resistivity data show that the small amount of the Mn impurities doped to the host system LnFeAsO$_{1-x}$F$_x$ (Ln=La and Nd; $x$=0.11) induces the significant increase of the low-$T$ resistivity. The result is contrasted with the $y$ dependence of the low-$T$ resistivity ($T>T_c$) observed for M=Co: For the Co doping, it first increases with increasing $y$, but the increasing rate is much smaller than for M=Mn. With further increasing Co concentration $y$ through ~0.05, the low-$T$ resistivity decreases and the system goes to the nonsuperconducting metallic phase at $y$~0.08 for Ln=La and at the $y$ value slightly larger than 0.10 for Ln=Nd. For M=Mn, the low-$T$ resistivity monotonically increases with $y$, indicating the significant effect of the electron localization. These behaviors of the resistivities are discussed in *4.1* to study the impurity effects on $T_c$ in relation to the symmetry of the order parameter.

The $T_c$ values are summarized in Figs. 3(a) and 3(b). From these figures, the values of the initial suppression rate $|dT_c/dy|$ for LaFe$_{1-y}$M$_y$AsO$_{1-x}$F$_x$ ($x$=0.11) are estimated to be ~2.5 K/% and 55 K/% for M=Co and Mn, respectively. Those for NdFe$_{1-y}$M$_y$AsO$_{1-x}$F$_x$ are ~2.9 K/% and 9 K/% for M=Co and Mn, respectively. It is remarkable that the $T_c$-suppression rates of Co and Mn impurities are quite different.

### 3.2 Thermoelectric powers and Hall coefficients of LnFe$_{1-y}$M$_y$AsO$_{1-x}$F$_x$ (Ln=La, Nd; M=Co, Mn; x=0.11)

The $T$ dependence of the thermoelectric powers $S$ of the samples of LaFe$_{1-y}$M$_y$AsO$_{1-x}$F$_x$ ($x$=0.11) are shown in Fig. 4(a) for M=Co and in Figs. 4(b) and 4(c) for M=Mn. Similar data of the Hall coefficient $R_H$ of LaFe$_{1-y}$Co$_y$AsO$_{1-x}$F$_x$ ($x$=0.11) are shown in Fig. 5(a), and those of NdFe$_{1-y}$M$_y$AsO$_{1-x}$F$_x$ are in Figs. 5(b) and 5(c) for M= Co and Mn, respectively. From the figures, we can see that both $S$ and $R_H$ exhibit anomalous $T$ dependences, in particular in the superconducting $y$ region, similar to those found in high $T_c$ Cu oxides,[21, 22] though the anomalous behaviors of $R_H$ of NdFe$_{1-y}$M$_y$AsO$_{1-x}$F$_x$ (M=Co and Mn) are less significant than that of LaFe$_{1-y}$Co$_y$AsO$_{1-x}$F$_x$. The results indicate that the system is magnetically active as in the superconducting Cu oxides.[23] Figures 6(a) and 6(b) show the thermoelectric powers $S$ and Hall coefficients $R_H$ of LnFe$_{1-y}$M$_y$AsO$_{1-x}$F$_x$ (Ln=La and Nd) at 100 K against ($x+y$) for M=Co and ($x-y$) for M=Mn, where the values on the horizontal axes correspond to the electron numbers of the systems measured from the nondoped system of LnFeAsO. (We show later that at least for the Co-doped system, the rigid band picture can be applied reasonably well.) Anomalous dips appear in the region of the above electron numbers between ~0.05 and ~0.2. The region essentially corresponds to the superconducting $x$ region [1] of LnFeAsO$_{1-x}$F$_x$.

### 3.3 Specific heats of LaFe$_{1-y}$M$_y$AsO$_{1-x}$F$_x$ (M=Co, Mn; x=0.11)

Figures 7(a) and 7(b) show the several examples of the specific heat($C$) data obtained for LaFe$_{1-y}$Co$_y$AsO$_{1-x}$F$_x$ ($x$=0.11). Similar plots are also shown for the samples of LaFe$_{1-y}$Mn$_y$AsO$_{1-x}$F$_x$ ($x$=0.11) in Fig. 7(c). In each figure, the $y$ values are attached to the corresponding data. In Fig. 7(a), we can see anomalies at temperatures $T_c$. From the anomalies, we roughly estimated the jump magnitudes $\Delta C$ of the specific heat at $T_c$, as shown in Fig. 8(a), for example. Then, the electronic specific heat coefficients γ were calculated by simply assuming the relation $\Delta C/T_c = 1.43\gamma$. For the nonsuperconducting samples, the γ values were roughly determined as the value at $T \to 0$ from the data shown in Figs. 7(b) and 7(c), because the linear relationship between $C/T$ and $T^2$ is not so apparent due to the magnetically active nature of the systems. (Only for the Co-doped sample with $y=0.7$ and the Mn-doped one with $y=0.05$, we estimated the γ values by neglecting the low-$T$ upturn of the specific heat observed as $T$ decreases.) The γ values thus obtained are plotted in Fig. 8(b) against $y(-y)$ for the Co(Mn) impurities (Again, the values of $0.11+y$ and $0.11-y$ correspond to the electron numbers in the the Co- and Mn-doped systems, respectively. In Fig. 8(c), the calculated data of the electronic density of states $N(E)$[24] is shown, for comparison, against the electron energy $E$, where the Fermi energy ($E_F$) of LaFeAsO is located at $E=0$ and $E_F$ of LaCoAsO is at the position indicated by the arrow. By comparing the characteristic structures of the calculated $N(E)$ and the observed γ, we find that the rigid band picture is essentially applicable to the system, as noted above.

*3.4 T dependence of the $^{75}$As-NMR $1/T_1$ of LaFe$_{1-y}$Co$_y$AsO$_{1-x}$F$_x$ (x=0.11)*

Figure 9 shows the $T$ dependence of the $^{75}$As-NMR longitudinal relaxation rates obtained here for several samples of LaFe$_{1-y}$Co$_y$AsO$_{1-x}$F$_x$ ($x=0.11$) with $T_c \geq 26$ K. As we reported previously,[12] the relation $1/T_1 \propto T^6$ can be observed in the $T$ region between $\sim 0.4T_c$ and $T_c$, which is contrasted with the relation $1/T_1 \propto T^{2.5-3.0}$ reported by many groups.[13-17] We note here, the former relation is observed for the samples, irrespective of whether the rather rapid decrease slightly above $T_c$ (from ~ 40 K) is observed or not with decreasing $T$.[12] The $T^6$-like behavior has also been observed in refs. 20 and 25 for Ba$_{1-x}$K$_x$Fe$_2$As$_2$ samples. We also note that $1/T_1$ observed in our study for BaFe$_{1.8}$Co$_{0.2}$As$_2$ (not shown) has almost identical $T$ dependence to that reported by Ning et al.[26] It exhibits the $T^{3-4}$-like behavior just below $T_c$.

To explain the $T^{2.5-3.0}$ dependence of $1/T_1$ by the $S_\pm$ symmetry, Parker et al.[18] introduced effects of the intermediate scattering centers, which induce the energy levels within the superconducting gap. Yashima et al.[201] also considered the $S_\pm$ symmetry of the superconducting order parameter with the gap anisotropy or the existence of the two different gaps on the two disconnected Fermi surfaces around Γ and M points as the intrinsic feature of the system. However, as we pointed out in our previous paper,[12] the former consideration cannot explain why the two distinct $T$ dependences have been observed even for a sample with almost equal $T_c$ value,[15] because $T_c$ values of samples with intermediate scatterers should be significantly lower for the $S_\pm$ symmetry than that those without such scattering centers. The latter idea cannot explain the existence of the two distinct $T$ dependences itself, because, if the anisotropy or two distinct gaps is(are) the intrinsic feature of the system, the $T^{2.5-3.0}$ dependences should be observed for all samples including those used in our measurements. We propose, in the next section, that the smaller exponent $n$ of the relation $1/T_1 \propto T^n$ is caused by the spatial inhomogeneity of $T_c$ or superconducting gap, by presenting additional data of $n$ obtained for various samples of LaFeAsO$_{1-x}$F$_x$ with different $x$ values or electron concentrations.

## 4. Discussion
*4.1 Impurity effects on $T_c$ and characteristics of the transport properties*

In *3.1-3.3*, effects of the Co and Mn dopings to LnFeAsO$_{1-x}$F$_x$ ($x=0.11$) on various quantities have been presented for Ln=La and Nd, where we have found anomalous $T$ dependence of the transport quantities. We have also found that the rigid band picture can be applied to the doped systems reasonably well. On the anomalous nature of the transport quantities, we have already pointed out[11] for LaFe$_{1-y}$Co$_y$AsO$_{1-x}$F$_x$ ($x=0.11$) that rather clear change exists in the transport behaviors with increasing $y$ or with increasing electron number $(x+y)=(0.11+y)$ doped into LaFeAsO from the superconducting to nonsuperconducting metallic region, and that the anomalous $T$ dependences of the thermoelectric powers $S$ and Hall coefficients $R_H$ reminiscent of those of high-$T_c$ Cu oxides[21, 22] are significant in the superconducting region of $(x+y)$ (we call this phase "anomalous metal"). In the present work, we have found that there exist similar anomalous $T$ dependences in all the systems shown in *3. 1* and *3. 2*. They are significant, in particular, in the region of the electron number, corresponding to the superconducting region (~0.05 ≤ $x$ ≤~0.2) of LaFeAsO$_{1-x}$F$_x$. This feature can be clearly seen as the anomalous dip structures of the $S$-$y$ and $R_H$-$y$ curves shown in Figs. 6(a) and 6(b) at 100 K for LaFe$_{1-y}$M$_y$AsO$_{1-x}$F$_x$ (M=Co and Mn; $x=0.11$). They indicate that the electronic state in the dip region [0.11≤ ($x+y$) ≤~0.2 for M=Co, and ~0.05 ≤ ($x-y$) for M=Mn] is different from those in the other regions of $y$. This point will be clarified after the discussion presented below.

The anomalous transport behaviors described above and a possible origin of the anomalous dips observed in Figs. 6(a) and 6(b) can be understood as described below. Here, we adopt the rigid band picture, which has been found, as already shown above, to be applicable to the present system. First, we consider the Co-doping, $(x+y) \geq 0.11$. In this case, the number of electrons donated by F and Co atoms into LnFeAsO is $(x+y)$, and these electrons bury the hole pockets around the Γ point in the reciprocal space, causing the disappearance (or reduction) of the hole-Fermi-surfaces around Γ. Due to this disappearance of the hole Fermi surfaces with heavy effective mass,[24] the strong magnetic fluctuations of this Fermi surface disappear, allowing the electrons around the M point to move without suffering from the scattering by this magnetic fluctuations. Judging from the data in Figs. 6(a) and 6(b), this change takes place at $(x+y)\sim 0.2$, and the transport properties becomes ordinary type ones as in the case of high-$T_c$ Cu oxides. (In high-$T_c$ Cu oxides, the anomalous transport behaviors diminish as the number of hole-carriers doped to the Mott insulating state increases and the magnetic fluctuation is weakened by the holes.[23]) This consideration is supported by the specific heat data shown in Fig. 8(b), where the minimum of the γ or $N(E)$ appears at $y\sim 0.1$ or $(x+y)\sim 0.2$ with increasing $y$ or $E$, indicating that the hole-Fermi-surfaces disappear at this concentration. The coincidence of this concentration with that of the upper edge of the dips of the curves in Figs. 6(a) and 6(b) suggests that for the dip structure, the existence of the hole-Fermi-surfaces is important. Because the superconductivity seems to occur within the region of the anomalous dips, we think that the existence of the hole-Fermi-surfaces is also important for the occurrence of the superconductivity.

To consider the case of Mn doping side, that is, for $(x-y) \leq 0.11$, we discuss, for a while, the data of the electrical resistivities $\rho$ in Figs. 10(a) and 10(b) obtained for LnFe$_{1-y}$M$_y$AsO$_{1-x}$F$_x$ (Ln=La and Nd; M=Co and Mn) at the temperatures indicated in the figures. In Fig. 10(a), data are intentionally included for samples of LaFe$_{1-y}$Co$_y$AsO$_{1-x}$F$_x$ in which the actual F concentrations are appreciably smaller than the nominal value (=0.11), and therefore their resistivities are larger than those of the samples without the reduction of $x$ from the nominal value. It is interesting that $T_c$ values of each series of samples in the figure are almost on a single line. [We note here the followings. The deviation of $x$ from the nominal value to the lower value side can be distinguished as the deviations of the resistivities and lattice parameters $c$ towards the larger value sides and the $T_c$ value towards the lower value side. This kinds of deviations cannot be explained by the change of $y$, because with increasing(decreasing) $y$ in the region $y < 0.05$, the resistivity increases(decreases) and $T_c$ and $c$ decrease(increase). The effects of the $x$ deviation are not so significant for NdFe$_{1-y}$M$_y$AsO$_{1-x}$F$_x$ (M=Co or Mn and $x$ =0.11).] We also included the data reported by Karkin et al.[27] for a sample whose lattice-defect density was controlled by successive heat treatments after the neutron irradiation without changing the carrier number density. From the figure, we find that the superconductivity disappears, roughly speaking, at the common value of the critical resistivity of ~3 mΩ·cm. (The absolute values of the resistivity of polycrystalline samples may not have so strict meaning, because they depend on the preparation conditions. For example, samples prepared *under the high-pressure syntheses often have smaller resistivities than those synthesized ordinary preparation processes*[28] possibly because of the difference of the grain boundary effects.) Here, in the analyses of the data plotted in Fig. 10(a), we estimate the in-plane resistivity expected for single crystals by multiplying the observed values of $\rho$ by a factor of ~1/4 to roughly remove grain-boundary- and anisotropy-effects, obtaining the critical in-plane resistivity of 0.75 mΩ·cm, which corresponds to the sheet resistances $R_\square$=8.6 kΩ (close to the value of 6.45 kΩ= $h/4e^2$, the inverse of the minimum metallic conductivity). The existence of the roughly common critical resistivity $\rho$, corresponding to the minimum metallic conductivity $4e^2/h$, suggests that the loss of the metallic nature (or the electron localization) is the origin of the $T_c$ vanishing, as has been known for the Cu-oxide system, too: For Bi$_2$Sr$_2$Y$_x$Ca$_{1-x}$Cu$_2$O$_8$, $T_c$ disappears when $R_\square$ exceeds 8 kΩ (~$h/4e^2$), even though the resistivity upturn is not so significant with decreasing $T$.[29] The idea is supported by the observation that $T_c$ vanishes at $\rho$~0.5 mΩ·cm[30] (or $R_\square$~7.7 kΩ ) for single crystals of Ba(Fe$_{1-x}$Co$_x$)$_2$As$_2$.

For LnFe$_{1-y}$Mn$_y$AsO$_{1-x}$F$_x$ ($x$=0.11; $x-y$<0.11), because the doping induces, as shown later, the significant increase of the resistivity, the zero $T_c$ is realized by the loss of the metallic nature. The anomalous transport behaviors also disappear along with this loss of the itinerant nature.

Now, the two $T_c$ vanishing mechanisms have been found to be relevant in the Co and Mn doped LnFeAsO$_{1-x}$F$_x$ ($x$=0.11) system: One is the electron localization, which appears when the sheet resistance $R_\square$ exceeds $h/4e^2$=6.45 kΩ, and another is the disappearance (or reduction) of the hole-Fermi-surfaces around the Γ point, which have strong magnetic fluctuations essentially important for the appearance of the "anomalous metallic phase" where the anomalous transport behaviors are significant. It is evident in Fig. 10(b) that the low temperature resistivity becomes very small as the system goes into the nonsuperconducting metallic phase upon the Co (or electron doping.) We emphasize that any experimental evidence has not been found that the pair breaking effect of nonmagnetic impurities expected for the $S_\pm$ symmetry superconductors has a primary role in the $T_c$ suppression.

Here, supposing that the pair breaking by the impurity scattering exists in the present system, we estimate the pair breaking parameter α. First, we show in Figs. 11(a)-11(d) the $y$ dependence of the residual resistivities of LnFe$_{1-y}$M$_y$AsO$_{1-x}$F$_x$ (Ln=La or Nd; M=Co or Mn; $x$=0.11), which were obtained by the extrapolation of the ρ-$T$ curve to $T$=0 from the $T$ region where the electron localization or the resistivity upturn with decreasing $T$ is not appreciable. From the figure, we find that the initial rate of the resistivity increase with $y$ is at least ~0.3 mΩ·cm/% for polycrystal samples of LaFe$_{1-y}$Co$_y$AsO$_{1-x}$F$_x$ ($x$=0.11). To estimate the impurity-scattering rate $1/\tau$ of the electrons, we consider that resistivities of single crystal samples are, as stated above, smaller than those of polycrystals by a factor of ~1/4, and other parameters are estimated as follows. For the estimation of the carrier number $n$ at $y$=0, we have adopted the $R_H$ value of $7 \times 10^{-3}$ cm$^3$/C at rather high temperature of 200 K to avoid the effect of the anomalous increase of $R_H$ with decreasing $T$ induced by the magnetically active nature of the system (on this point, see the discussion in ref. 24 for Cu oxide systems). Then, by the assumption that $R_H$ is determined at this temperature by the electron carriers with lighter mass than that in the hole bands, $n$~$0.9 \times 10^{21}$ /cm$^3$, consistent with that of the band calculation,[24] has been obtained. The $ab$ plane effective mass $m^*$ is considered to be equal to the free electron mass $m$, which can also be deduced approximately on the basis of the band calculation.[24] Then, using the relation $1/\rho = m^*/(ne^2\tau)$, we simply estimate the electron scattering time $\tau$, and the pair breaking parameter $\alpha = \hbar/(2\pi k_B T_{c0}\tau)$, which appears in the following well-known equation to determine the $T_c$ value for the $S_\pm$ symmetry in the presence of nonmagnetic impurities

$$\ln(T_{c0}/T_c) = \psi(1/2 + \alpha/2t) - \psi(1/2),$$

where $\psi(z)$ is the digamma function defined as $\psi(z) \equiv \ln\{d\Gamma(z)/dz/\Gamma(z)\}$, $T_{c0}$ is the superconducting $T_c$ of the nondoped system (~28 K for LaFeAsO$_{1-x}$F$_x$ ) and $t=T_c/T_{c0}$. Numerically, α is estimated from the residual resistivity shown above to be 0.84 at $y$=0.01. (Note that $T_c$ becomes zero at $\alpha \cong 0.28$).The initial slope $dT_c/dy$ can be given to be about -60K/%, which is much larger than the observed value of about -2.5 K/%.

For NdFe$_{1-y}$Co$_y$AsO$_{1-x}$F$_x$, the initial rate of the resistivity increase with $y$ is at least ~0.18 mΩ·cm/% for polycrystal samples. Using the $n$ value of $1.2\times10^{21}$/cm$^3$ similarly obtained to the case of LaFe$_{1-y}$Co$_y$AsO$_{1-x}$F$_x$, we can estimate the pair breaking parameter $\alpha$ {=$\hbar/(2\pi k_B T_{c0}\tau)$; $T_{c0}$=50 K in this case}, as for the LaFe$_{1-y}$Co$_y$AsO$_{1-x}$F$_x$ samples to be 0.39 at $y$=0.01. This result indicates that very small amount of Co (less than 1 %) completely suppresses the superconductivity, if the pair breaking effect by nonmagnetic scattering exists. We presume that detailed considerations of the coexistence of the hole- and electron-Fermi-surfaces may change the magnitude of the pair breaking by a certain factor. We may also have to consider other ambiguities introduced in the estimation of the parameters used in the above analyses. However, it is not easy to explain the observed small effect of the Co doping on $T_c$. Therefore, we think the pair breaking effect of the nonmagnetic impurities is not playing a primary role in the $T_c$ suppression. This conclusion can be also obtained from the data reported by Karkin et al.[27] for the neutron irradiated sample, where no change of the electron number takes place. (In their case, the 0.9 mΩ·cm change of the residual resistivity induces the $T_c$ change only less than 2 K.)

For the Mn doping, the electron localization becomes significant at low temperatures upon the very small amount of the doping, and the superconductivity disappears due to the loss of the itinerant nature. Figure 11(e) shows the schematic behavior of the residual resistivity against $(x+y)$ for LaFe$_{1-y}$Co$_y$AsO$_{1-x}$F$_x$. In the figure, the qualitative $(x+y)$ dependence of the scattering rates by the doped Co atoms, $1/\tau_{Co}$, and by the magnetic fluctuations, $1/\tau_m$, are also shown ($1/\tau_m$ approaches zero as $T$ goes to zero. Although the absolute value of $1/\tau_m$ does not have any meaning, the $(x+y)$ or $(x-y)$ dependence can be presumed from the NMR $1/T_1$ data[12, 31]). Even when $y$ is increased, the resistivity does not reach the critical value of $\rho_c$~3 mΩ·cm (or the critical sheet resistances $R_\square$=6.45 kΩ) for LaFe$_{1-y}$Co$_y$AsO$_{1-x}$F$_x$ (x=0.11) and superconductivity remains until the hole Fermi surface around the Γ point in the reciprocal space exhibits the significant reduction.

Onari and Kontani[32] have pointed out, by using the five band model, that the $T_c$ suppression by nonmagnetic impurities is expected to be significant. Recent experimental findings that various species of impurity atoms such as Rh, Ni, Ru and Ir[33-37] also induce the superconductivity, suggest that the suppression of the superconductivity by these impurities is not large. Considering these result, the insensitiveness of $T_c$ of the present Co-doped system should be considered seriously in arguing the superconducting symmetry, though above arguments seem to conflict with the data of neutron scattering, which suggest the existence of the so-called resonance peak expected for the $S_\pm$ symmetry.[38, 39] We also note here that there has been announced[40] quite recently that the ~1.5 % Zn doping to the LaFeAsO$_{0.85}$ system prepared *by the high pressure syntheses* completely suppresses $T_c$, where the increase of the residual resistivity by ~0.9 mΩ·cm was observed with a slight resistivity upturn with decreasing $T$ at low $T$. (The absolute value is ~1.2 mΩ·cm.) Detailed studies are necessary on that system, too, to clarify which of the localization- or pair-breaking mechanism can explain the results.

*4.2 On the two distinct $^{75}$As-NMR $1/1/T_1$ of LaFe$_{1-y}$Co$_y$AsO$_{1-x}$F$_x$ (x=0.11)*

As stated in *3.2*, the two distinct $T$ dependences have been observed for the $^{75}$As-NMR $1/T_1$ of LaFeAsO$_{1-x}$F$_x$ (x=0.11). Although the explanations of the relation $1/T_1 \propto T^{2.5-3.0}$ has been proposed by considering the $S_\pm$-symmetry[18, 20], there exist difficulties to understand these data consistently. To solve these difficulties, we first show Fig. 12, where the maximum $T_c$ values of LnFeAsO$_{1-\delta}$ reported previously[41] are shown against the lattice parameter $c$ by open circles for various rare earth elements Ln. The $T_c$ values of LaFeAsO$_{1-x}$F$_x$ and NdFeAsO$_{1-x}$F$_x$ prepared here starting from the initial $x$ value of 0.11 are also shown by solid circles for many samples from different batches. Insets shows the data for Ln= La and Nd with the expanded horizontal scales. We can see that $T_c$ depends on the lattice parameter $c$ very sensitively, that is, for the larger $c$, which corresponds, as described in *4. 1*, to the smaller F concentration,[42] the $T_c$ value becomes smaller. Detailed structure analyses by Rietveld method have also given the same results. We think that this change of the F concentration is the origin of the irregular $y$ dependence of $T_c$ observed previously for LaFe$_{1-y}$Co$_y$AsO$_{1-x}$F$_x$ (x=0.11).[10, 11] This kind of irregular $y$ dependence is not so significant for NdFe$_{1-y}$Co$_y$AsO$_{1-x}$F$_x$ (x=0.11) as compared with that for LaFe$_{1-y}$Co$_y$AsO$_{1-x}$F$_x$.

The above considerations present a clue to answer the puzzle of the $1/T_1$, that is, the spatial inhomogeneity or irregularity of the superconducting $T_c$ or the order parameter may affect the $T$ dependence of $1/T_1$. Because the effect is not an intrinsic one, it depends on samples used in the measurements, explaining the existence of various reported $T$ dependences.

In Figs. 13(a) and 13(b), results of the trial calculations of $1/T_1$ with (red line) and without (blue line) the distribution of the order parameter, are shown, where the inhomogeneity or irregularity is considered as the distribution of the quasi particle density of states. In the former figure, the coherence factor for an isotropic order parameter is considered, and in the latter, it is not considered, and therefore can be used for the $S_\pm$ symmetry. The BCS type $T$ dependence is used for the superconducting order parameter $\Delta$ and its zero $T$ value is $\Delta_0$. In Fig. 13(a), following parameters are used to obtain the blue line: $2\Delta_0/k_B T_c$=8. The Lorentzian type broadening $\Gamma$=0.24$\Delta_0$ of the quasi particle energy $\varepsilon$ is simply introduced with the cut-off of the quasi particle density of states $N(\varepsilon)$ at the energy of $0.9 k_B T_c$, below which $N(\varepsilon)$ is zero. To obtain the red line of Fig. 13(a), the coexistence of three values of the parameters $2\Delta_0/k_B T_c$=8, 6.5 and 5 are considered with equal weights, and their broadenings are all $\Gamma$=0.24$\Delta_0$ and the cut-off energy is $0.3 k_B T_c$. In Fig. 13(b), the parameters for the blue line are as follows: $2\Delta_0/k_B T_c$=6, $\Gamma$=0.07$\Delta_0$ and no cut-off has been introduced. To obtain the red line of Fig. 13(b), the coexistence of three values of the parameters $2\Delta_0/k_B T_c$=6, 4 and 2 with equal weights are assumed, and their broadenings are all 0.12$\Delta_0$ and the cut-off energy is $0.25 k_B T_c$. The rather broad distribution of $\Delta_0$ or $T_c$ is not unrealistic for samples, because, as shown in Fig. 12, $T_c$ is very sensitively depends on $c$ or $x$. The inhomogeneity of $\Delta_0$ or $T_c$ is considered to be more significant for samples, with the $T_c$'s smaller than the maximum one, because $|dT_c/dx|$ is larger. As we can find from the Figs. 13(a) and 13(b), the introduction of the simple distribution of the order parameter $\Delta$ can explain the difference rather well. The above results are not very different from those expected for anisotropy considerations of the order parameter in the sense that the broadening of $\Delta$ is introduced. However, our model states that the $T^6$-like behavior of $1/T_1$ observed in the rather wide $T$ region of the superconducting phase from immediately below $T_c$ is the one expected for samples without spatial inhomogeneity. It is contrasted with the models to explain the relation $1/T_1 \propto T^{2.5-3.0}$ by the effects of the intermediate scatterers[18] or by the intrinsic distribution of the order parameters.[20] We add here that the detailed Rietveld analyses of our samples used in the NMR studies have found that the structural parameters do not indicate any deviation from the published data, indicating that there is no reason to conceive the difference between the structural parameters as the origin of the different behaviors of $1/T_1$.

The above arguments are supported by the following data. In Figs. 14(a) and 14(b), the exponent $n$ obtained by fitting the relation $1/T_1 \propto T^n$ (~$0.4T_c < T < T_c$) to the observed data of LaFeAsO$_{1-x}$F$_x$ are plotted against (a)$T_c$ and (b)$1/T_1T$ at 50 K, with $x$ or the electron number doped to LaFeAsO being the internal parameter. These data directly indicate that the exponent $n$ depend on $x$. In the figure, our own samples are shown by the solid circles and the $x$ values are attached to the corresponding data points. Other data points except the open square correspond to the results reported by other group[14, 15, 17, 31, 43] (the open square indicates the $n$ value of a sample of LaFeAsO$_{1-\delta}$.[15]) The $1/T_1$-$T$ data obtained for $x$=0.15 are shown, for example, in Fig. 14(c). Because $1/T_1T$ monotonically depends on the doped electron number,[31] we simply expect that the data are on a single line in Fig. 14(b). Actually, our data of $n$ are on a single line (solid one). However, the data reported by other groups are found on a different line (dotted one). As for the plot against $T_c$, the data points except the open square are on a single line (gray line), even though $T_c$ has the dome-shape-dependence on $x$ or the electron number doped to LaFeAsO. The largest $n$ is found to correspond to the highest $T_c$ or optimal $x$. It is consistent with the above consideration: At the maximum $T_c$ point, because $|dT_c/dx|$ or $|d\Delta_0/dx|$ has the smallest value, and the distribution width of $T_c$ or $\Delta_0$ is expected to be the smallest. Then, considering that for the larger distribution width, $n$ is, as shown above, smaller, we can explain the difference among the exponents $n$. We can exclude the difference of $2\Delta_0/k_BT_c$ among the samples, as the origin of the observed difference of $n$, because the samples with equal $T_c$ values have different $n$ values and because those with equal values of $1/T_1T$ (or equal values of $x$) have different $n$ values.

Finally, we note that our NQR spectra obtained for the sample with the maximum $T_c$ (~28 K) do not show any significant anomaly to be considered as the origin of the $T^6$ dependence of $1/T_1$ different from the $T^{2.5-3.0}$ dependence reported by other groups. Based on the above results, we stress here that that the relation $1/T_1 \propto T^{2.5-3.0}$ cannot be, at least at present, the experimental evidence for the $S_\pm$ symmetry of the superconductivity of this system.

## 5. Summary

Results of the transport studies and data of the specific heat coefficients $\gamma$ of LnFe$_{1-y}$M$_y$AsO$_{1-x}$F$_x$ (Ln=La and Nd; M=Co and Mn; $x$=0.11) have been presented. The results of the measurements of the $^{75}$As-NMR $1/T_1$ of LaFe$_{1-y}$Co$_y$AsO$_{1-x}$F$_x$ have also been presented. The doped Co atoms are nonmagnetic, and the $T_c$-suppression rate $|dT_c/dx|$ by the Co atoms has been found to be too small to understand by the pair breaking effect. It throws a serious doubt whether the $S_\pm$ symmetry is realized in the system. For the present systems, the two mechanisms are relevant to the $T_c$-suppression, instead of the pair breaking by the doped impurities: One is the electron localization and another is the diminishing hole-Fermi-surfaces around the $\Gamma$ point in the reciprocal space.

On the two distinct $T$ dependences of the NMR longitudinal relaxation rate $1/T_1$ of LaFeAsO$_{1-x}$F$_x$ ($x$=0.11), we have proposed that the irregular distribution of the order parameter within the samples is a possible origin of the latter behavior, and that the $T^6$-like dependence is the one without the extrinsic spatial inhomogeneity of the order parameter. At least, it is difficult to consider the $T^{2.5-3.0}$-like as the firm experimental evidence for the $S_\pm$ symmetry of $\Delta$.

Acknowledgments – The authors thank Prof. H. Kontani for fruitful discussion. The work is supported by Grants-in-Aid for Scientific Research from the Japan Society for the Promotion of Science (JSPS), Grants-in-Aid on Priority Area from the Ministry of Education, Culture, Sports, Science and Technology and JST, TRIP.

Figure captions

Fig. 1 In (a), the lattice parameters $a$ and $c$ of LaFe$_{1-y}$M$_y$AsO$_{1-x}$F$_x$ ($x$=0.11) system plotted against $y$ and $-y$ values for M= Co and M=Mn, respectively. In (b) and (c), the electrical resistivities $\rho$ of the Mn- and Co-doped samples, respectively, are shown against $T$, where the $y$ value of each sample is attached to the corresponding data.

Fig. 2 Similar figures to Figs. 1(a)-1(c) are shown for NdFe$_{1-y}$M$_y$AsO$_{1-x}$F$_x$ (M=Co and/or Mn and $x$ =0.11).

Fig. 3 $T_c$ values are summarized for RFe$_{1-y}$M$_y$AsO$_{0.89}$F$_{0.11}$ (R=La and Nd; M=Co and Mn). They are plotted against $y$ and $-y$ for M=Co and M=Mn, respectively.

Fig. 4 $T$ dependence of the thermoelectric powers $S$ of the samples of LaFe$_{1-y}$M$_y$AsO$_{1-x}$F$_x$ ($x$=0.11) are shown for M=Co(a) and Mn(b). In (c), the selected data in (b) are shown with the different vertical scale.

Fig. 5 Hall coefficients $R_H$ of LaFe$_{1-y}$Co$_y$AsO$_{1-x}$F$_x$(a), and NdFe$_{1-y}$M$_y$AsO$_{1-x}$F$_x$ with M=Co(b) and Mn(c) are shown against $T$. The $y$ values are attached to the corresponding data.

Fig. 6 Thermoelectric powers $S$(a) and the Hall coefficients $R_H$(b) of LnFe$_{1-y}$M$_y$AsO$_{1-x}$F$_x$ (Ln=La and Nd; M=Co and Mn; $x$=0.11) are shown against $x+y$ and $x-y$ at 100 K for M=Co and Mn, respectively. The dip structures are obvious in both panels, showing the anomalous nature of the transport properties of the system in the narrow region of the carrier concentration.

Fig. 7 (a) Results of the specific heat measurements are shown for various values of the superconducting samples of LaFe$_{1-y}$Co$_y$AsO$_{1-x}$F$_x$ ($x$=0.11) against $T$. Anomalies can be observed at $T_c$. The $y$ values are attached to the corresponding data. (b) Data of $C/T$ are shown against $T^2$ for the selected values of $y$ of the nonsuperconducting metallic samples of LaFe$_{1-y}$Co$_y$AsO$_{1-x}$F$_x$ ($x$=0.11). The $y$ values are attached to the corresponding data. (c) Similar plots $C/T$ are shown for the samples of LaFe$_{1-y}$Mn$_y$AsO$_{1-x}$F$_x$ ($x$=0.11). The $y$ values are attached to the corresponding data.

Fig. 8 (a) The jump magnitudes of the electronic specific heat divided by $T$, $\Delta C/T$ of LaFe$_{1-y}$Co$_y$AsO$_{1-x}$F$_x$ at $T_c$ were estimated for the superconducting samples as shown in the figure, for example. (b) The electronic specific heat coefficients $\gamma$ of LaFe$_{1-y}$M$_y$AsO$_{1-x}$F$_x$ are shown against $y$ and $-y$ for M=Co and Mn, respectively. The $\gamma$ values were roughly estimated for the superconducting samples by simply assuming the BCS relation $\Delta C/T_c$=1.43$\gamma$. Those of the nonsuperconducting metallic samples were determined as the values at $T\rightarrow 0$. For M=Co and $y$=0.7, and for M=Mn and $y$=0.05,.the extrapolated values from the $T$ region above 50 K is used. (c) The electronic density of states calculated for LaFeAsO[23] is shown as a function of the electron energy. The arrow indicates the Fermi level of LaCoAsO. From these figures, we can see that the rigid band picture works reasonably well.

Fig. 9 $T$ dependence of the $^{75}$As-NMR longitudinal relaxation rates of several samples of LaFe$_{1-y}$Co$_y$AsO$_{1-x}$F$_x$ ($x$=0.11). As we reported previously,[12] the relation $1/T_1 \propto T^6$ can be observed in the $T$ region between ~0.4$T_c$ and $T_c$. Inset shows the data with the linear scales.

Fig. 10 (a) $T_c$ values are plotted against the electrical resistivities $\rho$ for LaFe$_{1-y}$M$_y$AsO$_{1-x}$F$_x$ and NdFe$_{1-y}$M$_y$AsO$_{1-x}$F$_x$ (M=Co and Mn) at the temperatures indicated in the figure.. Data for samples of LaFe$_{1-y}$Co$_y$AsO$_{1-x}$F$_x$ with the F concentration $x$ smaller than the nominal value (=0.11) are included. The data of Karkin $et$ $al$.[27] are also shown. (b) $T_c$ values are plotted against the resistivities obtained at the temperatures indicated in the figure, for the samples of LaFe$_{1-y}$Co$_y$AsO$_{1-x}$F$_x$ and NdFe$_{1-y}$Co$_y$AsO$_{1-x}$F$_x$ in the region near the boundary between the superconducting ($y<y_c$) and the nonsuperconducting metallic ($y>y_c$) phases. The $y$ values are attached to the data points.

Fig. 11 In (a)-(d), the residual resistivities $\rho$ of LnFe$_{1-y}$M$_y$AsO$_{1-x}$F$_x$ (Ln=La and Nd; M= Co and Mn; $x$=0.11), are plotted against $y$. They are obtained by extrapolating the $\rho$-$T$ curve to $T$=0 from the $T$ region where the resistivity upturn with decreasing $T$ is not appreciable. The initial increasing rate of $\rho$ with $y$ can be estimated by using the broken lines for the Co doping. (e) Schematic behavior of the residual resistivity is shown for LaFe$_{1-y}$Co$_y$AsO$_{1-x}$F$_x$ against ($x+y$). In the figure, the ($x+y$) dependence of the scattering rates, $1/\tau_{Co}$ by the doped Co atoms and by the magnetic fluctuations, $1/\tau_m$ are shown schematically, where their relative magnitudes do not have a meaning. The superconducting region of LaFe$_{1-y}$Co$_y$AsO$_{1-x}$F$_x$ is shaded. The horizontal axis region <0.11 corresponds to the F concentration $x$ smaller than the nominal value, or corresponds to ($x$-$y$) used for the Mn doping.

Fig. 12 Maximum $T_c$ values of LnFeAsO$_{1-\delta}$ obtained with changing $\delta$[32] are shown by open circles for various rare earth elements Ln. The $T_c$ values of LaFeAsO$_{1-x}$F$_x$ and NdFeAsO$_{1-x}$F$_x$ obtained for the samples prepared here starting from the nominal $x$ value of 0.11 are also shown by solid circles for many samples from different batches. They are very sensitive, in particular, for LaFeAsO$_{1-x}$F$_x$, to the lattice parameter $c$. Insets show the data with the enlarged horizontal axes.

Fig. 13 $1/T_1$-$T$ curves obtained by fitting calculated results to the data for our sample of LaFeAsO$_{0.89}$F$_{0.11}$ are shown by the blue solid lines.

The red lines were obtained by fitting to the curve of $1/T_1 \propto T^3$. The calculations have been carried out by the assumption that the distribution of the order parameter exists (red line) and does not exist(blue line). In (a), the coherence factor for an isotropic order parameter is considered, and in (b), it is not considered. The BCS type $T$ dependence is used for the superconducting order parameter. Detailed parameters are in the text.

Fig. 14 The values of the exponent $n$ obtained from the relation $1/T_1 \propto T^n$ are shown against (a) $T_c$ and (b) $1/T_1T$ at 50 K. In (c), $1/T_1$ of the sample of LaFeAsO$_{1-x}$F$_x$ with $x$=0.15 are shown against $T$ in the logarithmic scales.

Fig. 2

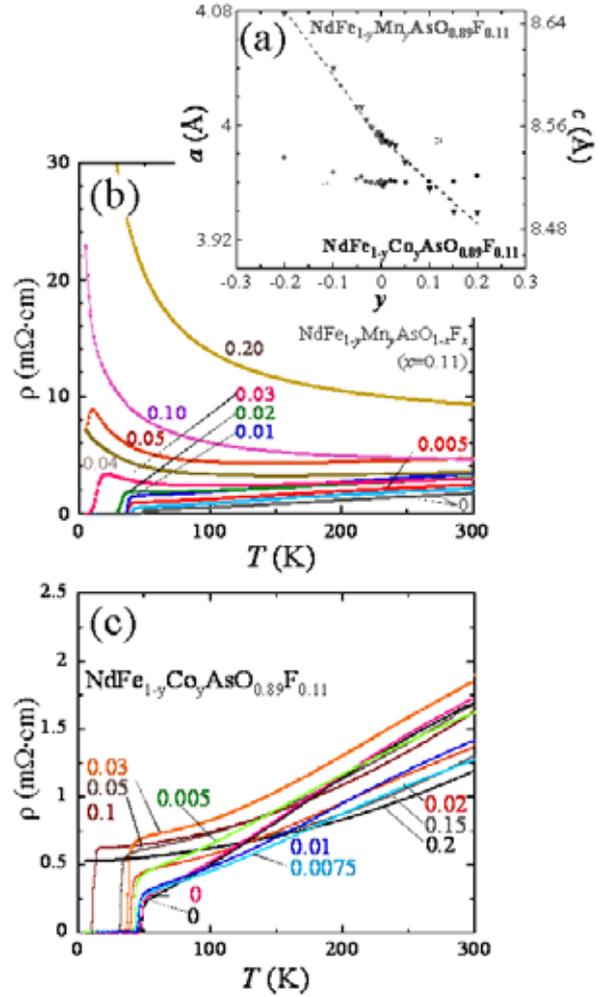

Fig.1

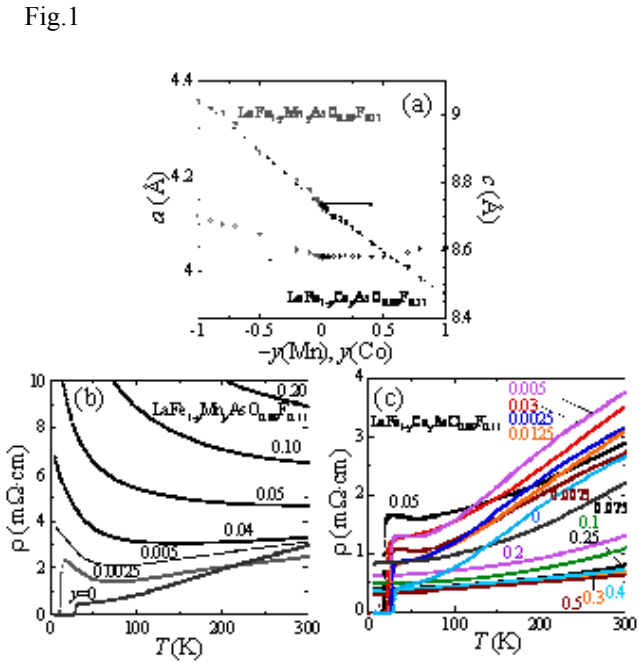

Fig. 3

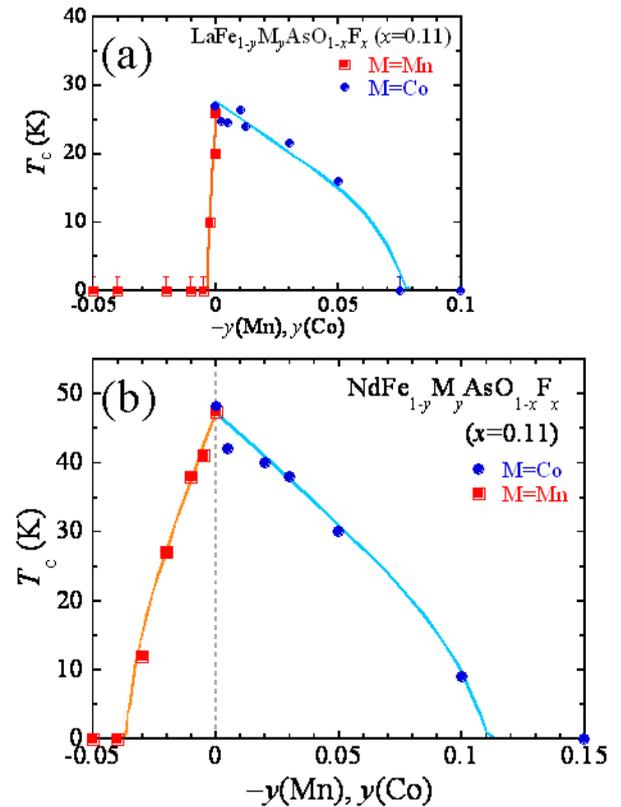

Fig. 4

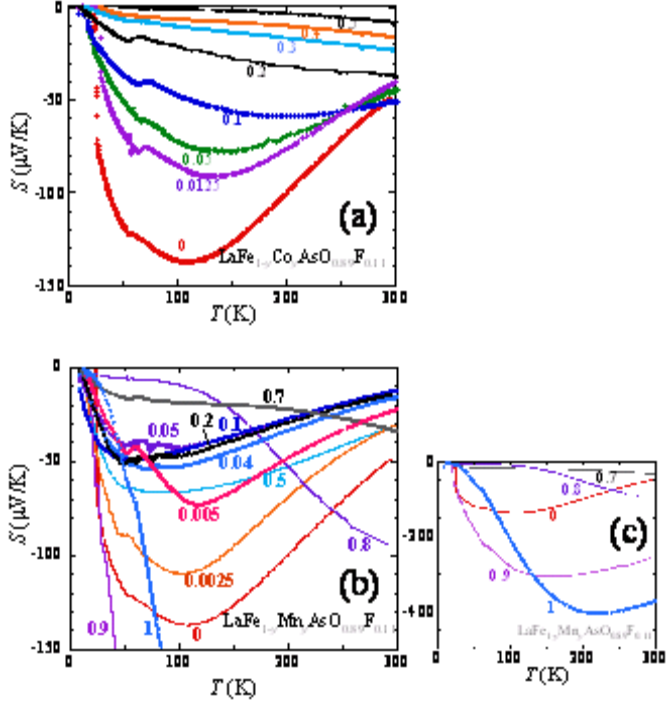

Fig. 5

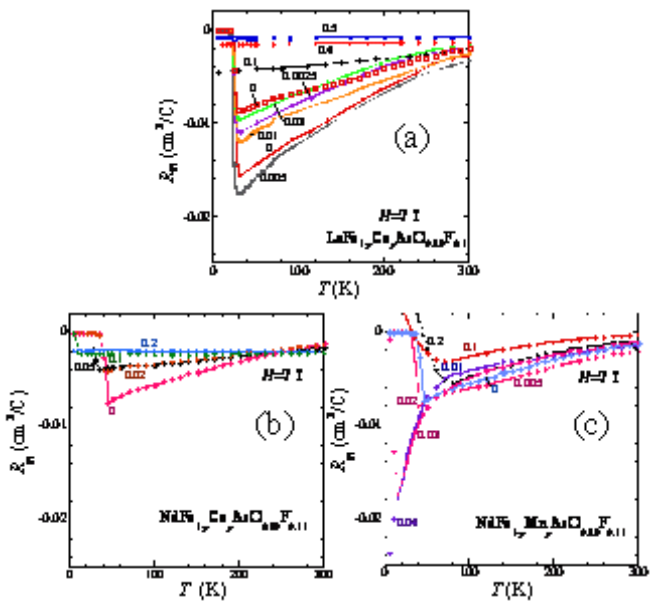

Fig. 6

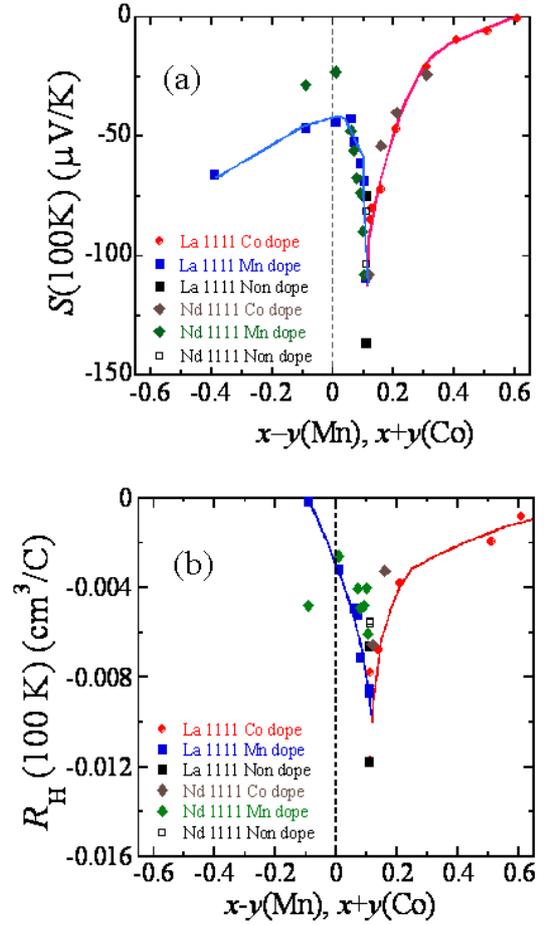

Fig. 7

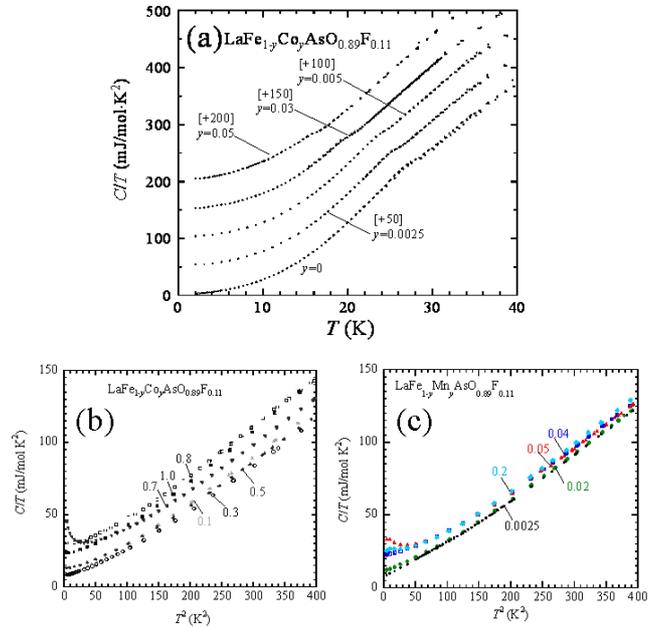

Fig. 8

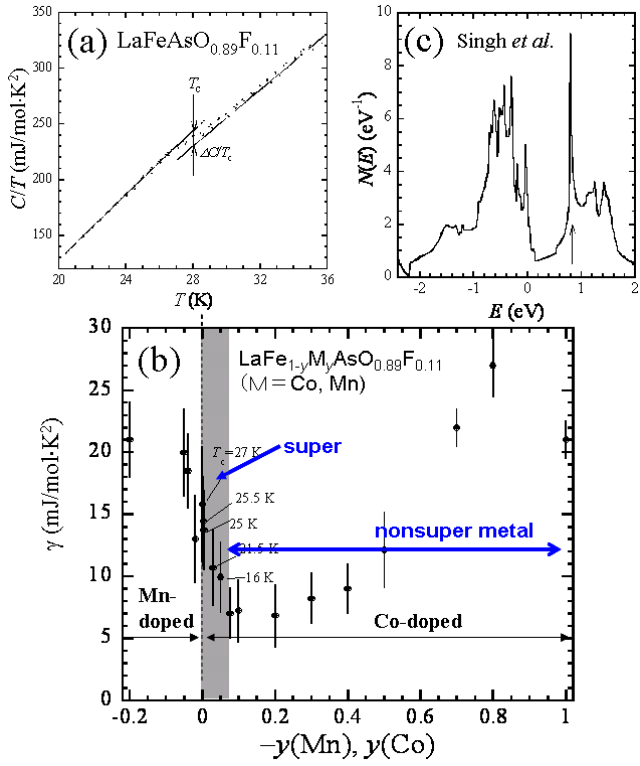

Fig. 9

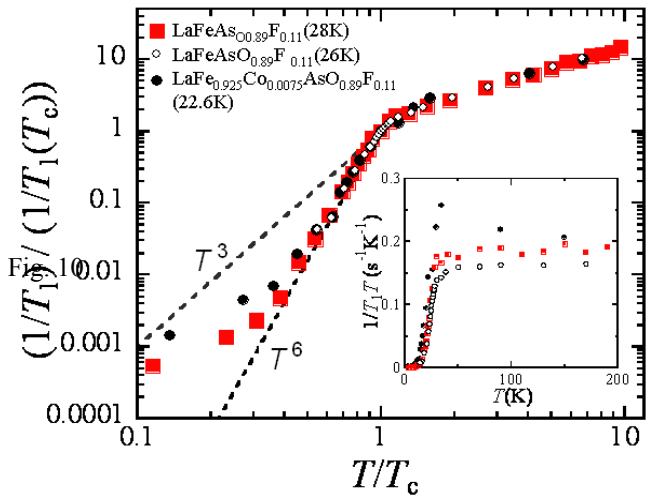

Fig. 10

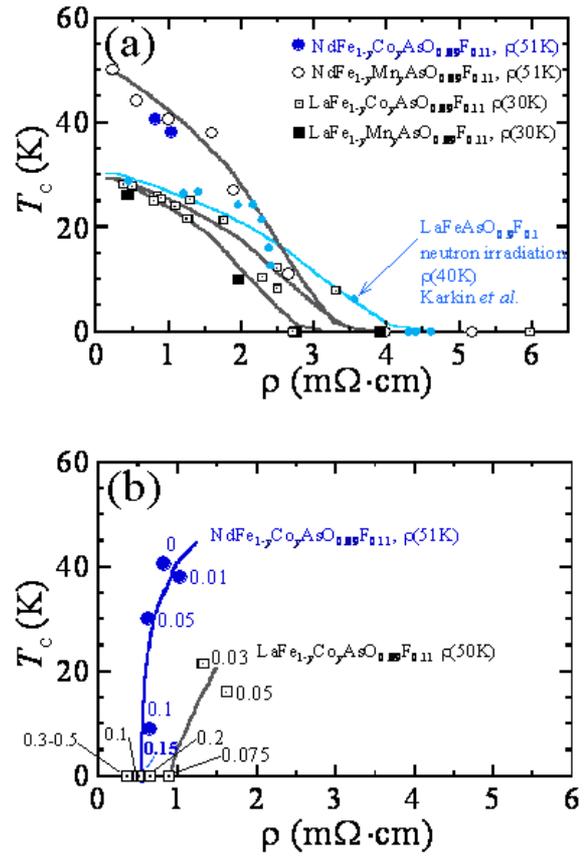

Fig. 11

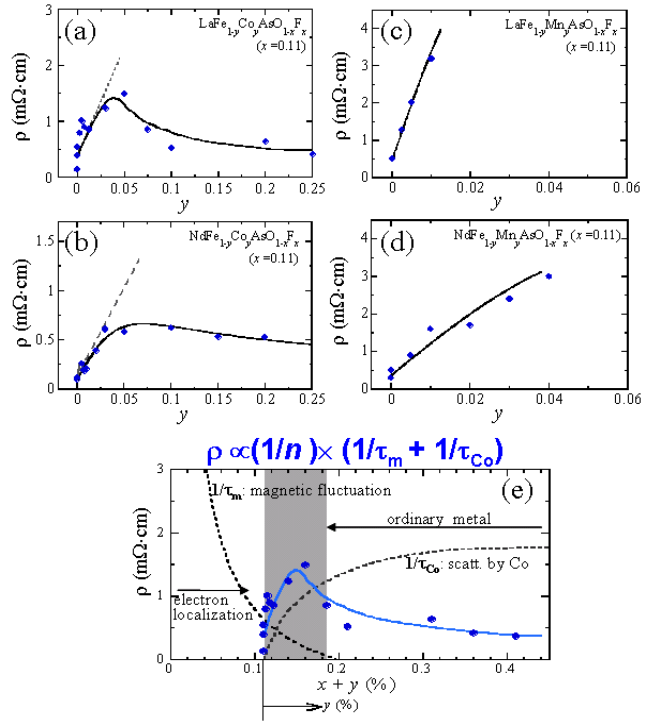

Fig. 12

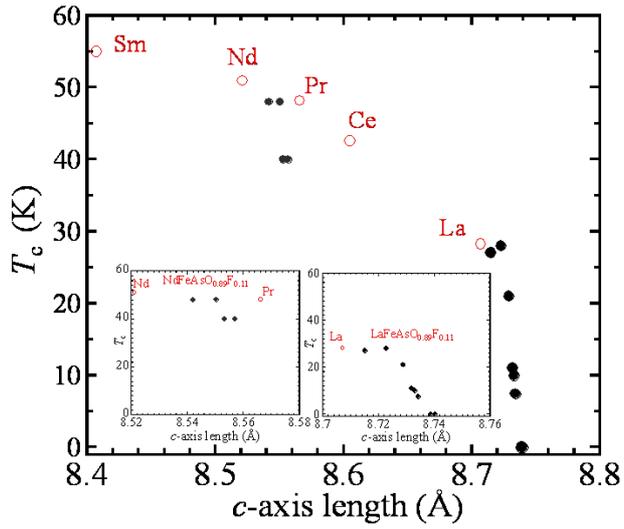

Fig. 14

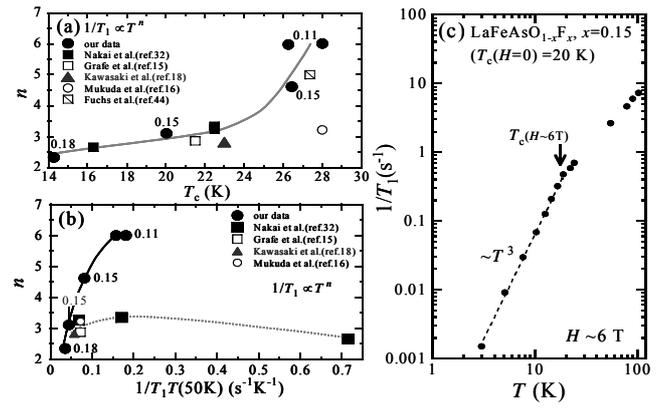

Fig. 13

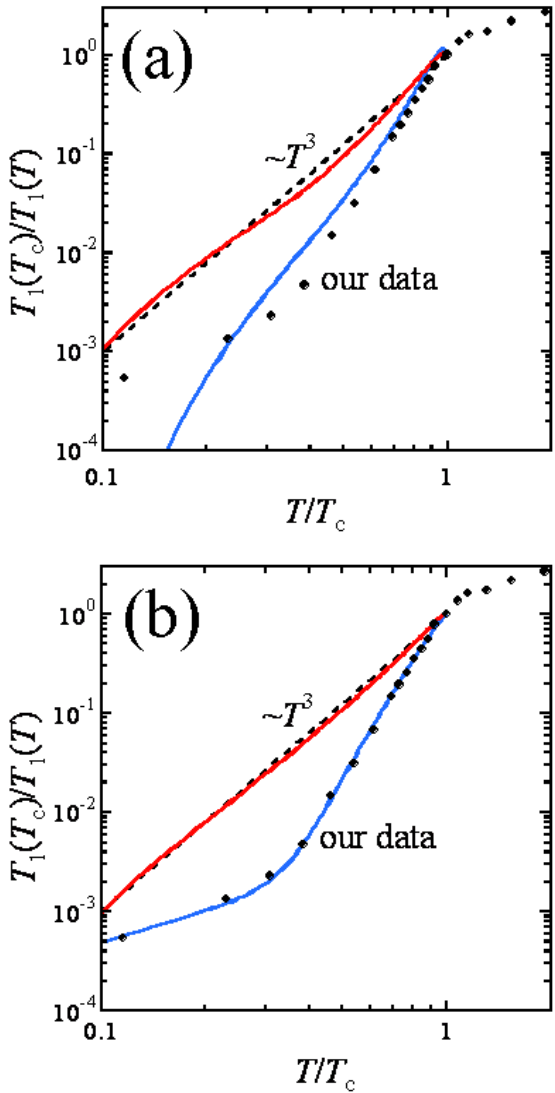